%% file: paper.tex
\documentclass[conference,a4paper]{IEEEtran}
\usepackage[underline=true]{pgf-umlsd}
\usepackage{subcaption}
\usepackage[mode=text]{siunitx}
\usepackage[mode=buildnew,subpreambles]{standalone}
\usepackage{hyperref}
\usepackage[capitalize,noabbrev]{cleveref}
\usepackage[sorting=none,maxbibnames=99]{biblatex}
\newcommand{\system}[0]{GLIDS}
\newcommand{\systemfull}[0]{Global Latency Information Dissemination System}
\newcommand{\mysubsubsection}[1]{\subsubsection{\textbf{#1}}}

\title{\system{}: A \systemfull{}}

\author{\IEEEauthorblockN{Cyrill Krähenbühl, Seyedali Tabaeiaghdaei, Simon Scherrer, Matthias Frei, Adrian Perrig}
\IEEEauthorblockA{\textit{Department of Computer Science, ETH Zürich, Switzerland}}}

\begin{document}

\maketitle

\begin{abstract}
  A recent advance in networking is the deployment of path-aware multipath network architectures, where network endpoints are given multiple network paths to send their data on.
  In this work, we tackle the challenge of selecting paths for latency-sensitive applications.
  Even today's path-aware networks, which are much smaller than the current Internet, already offer dozens and in several cases over a hundred paths to a given destination, making it impractical to measure all path latencies to find the lowest latency path.
  Furthermore, for short flows, performing latency measurements may not provide benefits as the flow may finish before completing the measurements.
  To overcome these issues, we argue that endpoints should be provided with a latency estimate \emph{before} sending any packets, enabling latency-aware path choice for the first packet sent.
  As we cannot predict the end-to-end latency due to dynamically changing queuing delays, we measure and disseminate the \emph{propagation} latency, enabling novel use cases and solving concrete problems in current network protocols.
  We present the \systemfull{} (\system{}), which is a step toward global latency transparency through the dissemination of propagation latency information.
\end{abstract}

\section{Introduction}\label{sec:introduction}\label{sec:intro}
\input{intro}

\section{Background}\label{sec:background}
\input{background}

\section{Motivation and Challenges}\label{sec:prediction}
\input{prediction}

\section{System Design}\label{sec:system-design}
\input{system-overview}

\section{Implementation and Evaluation}\label{sec:evaluation}
\input{evaluation}

\section{Discussion}\label{sec:discussion}
\input{discussion}

\section{Related Work}\label{sec:related}
\input{related}

\section{Conclusion}\label{sec:conclusion}\label{sec:concl}
\input{conclusion}

\section{Acknowledgments}\label{sec:acknowledgements}\label{sec:ack}
\input{ack}

\printbibliography

\appendix
\input{appendix}

\end{document}

%% file: intro.tex
With the deployment of path-aware network (PAN) architectures such as SCION~\cite{Kraehenbuehl2021}, new challenges and opportunities arise for latency-sensitive applications such as video conferencing, online gaming, holographic communication, or the tactile Internet~\cite{itu-network-requirements-2030}.
In contrast to today's Internet, which only offers a single, typically cost-optimized, path, a PAN can offer multiple path options, providing advantages to latency-sensitive applications since endpoints can select the shortest-latency path.
In the current deployment of SCION, we witness that a large number of distinct paths are available, with many destinations having over 100 different paths.
With the steady expansion of the commercial SCION network, we expect this number to further increase.

For latency-sensitive applications, a research challenge thus emerges on which path to use.
The path length in terms of AS hops is unfortunately a weak predictor of the end-to-end latency~\cite{Fei1998}.
Path probing with active measurements would waste the benefit in the case of dozens, or even hundreds, of available paths.
In particular, single-packet request-response protocols, such as DNS, would not permit any probing for achieving low-latency operation.

In this work, we explore an approach that adds \textit{propagation} latency information to PAN path information, enabling an endpoint to compute an estimate for the end-to-end propagation latency.
The reasons for disseminating the propagation latency instead of attempting to predict the experienced end-to-end latency, which also comprises the transmission, processing, and queuing latency, are as follows.
Transmission and processing latencies are typically negligible compared to propagation and queuing, and often exhibit little variance.
Since the amount and nature of cross traffic is usually unpredictable, an accurate estimation of the queuing latency is not possible in most cases.
On the other hand, propagation latency is a useful metric, as it enables the computation of the experienced queuing latency, is typically static, and can be measured without relying on a model.
Finally, as we will outline below, for use cases such as efficient path probing and improving the fairness of congestion control algorithms, knowledge of the propagation latency provides significant benefits.

We propose the \systemfull{} (\system{}) for estimating propagation latency information of end-to-end paths in an inter-domain setting.
We study the research problem of how to achieve high accuracy for latency estimates while minimizing the network overhead.
Furthermore, since \system{} is based on SCION, where partial paths in the form of path segments are combined into end-to-end paths, we ensure that end-to-end propagation latency estimation is possible for all path segment combinations.

One goal of \system{} is to find the lowest latency path in a multipath network with a large number paths.
For long-lived latency-critical flows, a sender will (continuously) probe available paths and switch to the lowest end-to-end latency path available.
Knowledge of the propagation latency gained through \system{} provides the sender with a clear order and cutoff latency for path probing instead of relying on heuristics which may not find the lowest experienced latency path.

For short-lived flows that finish in a single round trip, e.g., consisting of a single request and response packet pair or where all data fits into the initial congestion window, performing latency measurements to decide where to send a packet may take as long as, or even longer than, the flow itself and thus negate any benefits of the measurement.
Even for longer-lived flows, knowledge of the propagation latency gained from \system{} is valuable to make an informed path selection when sending the \textit{first} packet.

Another use of \system{} is to improve delay-based congestion control algorithms (CCAs), which adjust a sender's congestion window based on the measured latency inflation as a signal of queuing on intermediate nodes.
BBR is a delay-based CCA which has shown to be unfair to competing loss-based CUBIC CCA flows, due to overestimating the propagation latency~\cite{hock2017experimental,mishra2022we}.
By providing a propagation latency estimate, \system{} enables BBR to infer the current latency inflation and overcome this unfairness.

Compared to the current Internet, the key ingredient enabling \system{} in the SCION path-aware Internet architecture is the stability of disseminated paths.
\system{} leverages this stability to directly use the latency information disseminated through the control-plane for path selection and thus provides propagation latency transparency.
The main contributions of this work are:

\begin{itemize}
\item Design of the scalable system \system{} for estimating propagation latency of end-to-end paths in SCION.
\item Evaluating \system{} through the SCIONLab testbed, emulation in a Mininet-based BBR testbed, and simulations on real-world topologies.
\end{itemize}

%% file: background.tex
This section briefly introduces the main components of the next-generation Internet architecture SCION~\cite{Chuat2022a} that are relevant for this work.
SCION enables path transparency, i.e., endpoints know which autonomous systems (ASes) and border routers their traffic traverses, and path control, i.e., endpoints can influence which forwarding path is taken.

\subsubsection{Control Plane}

\begin{figure}[t]
  \centering
  \includegraphics[width=\linewidth]{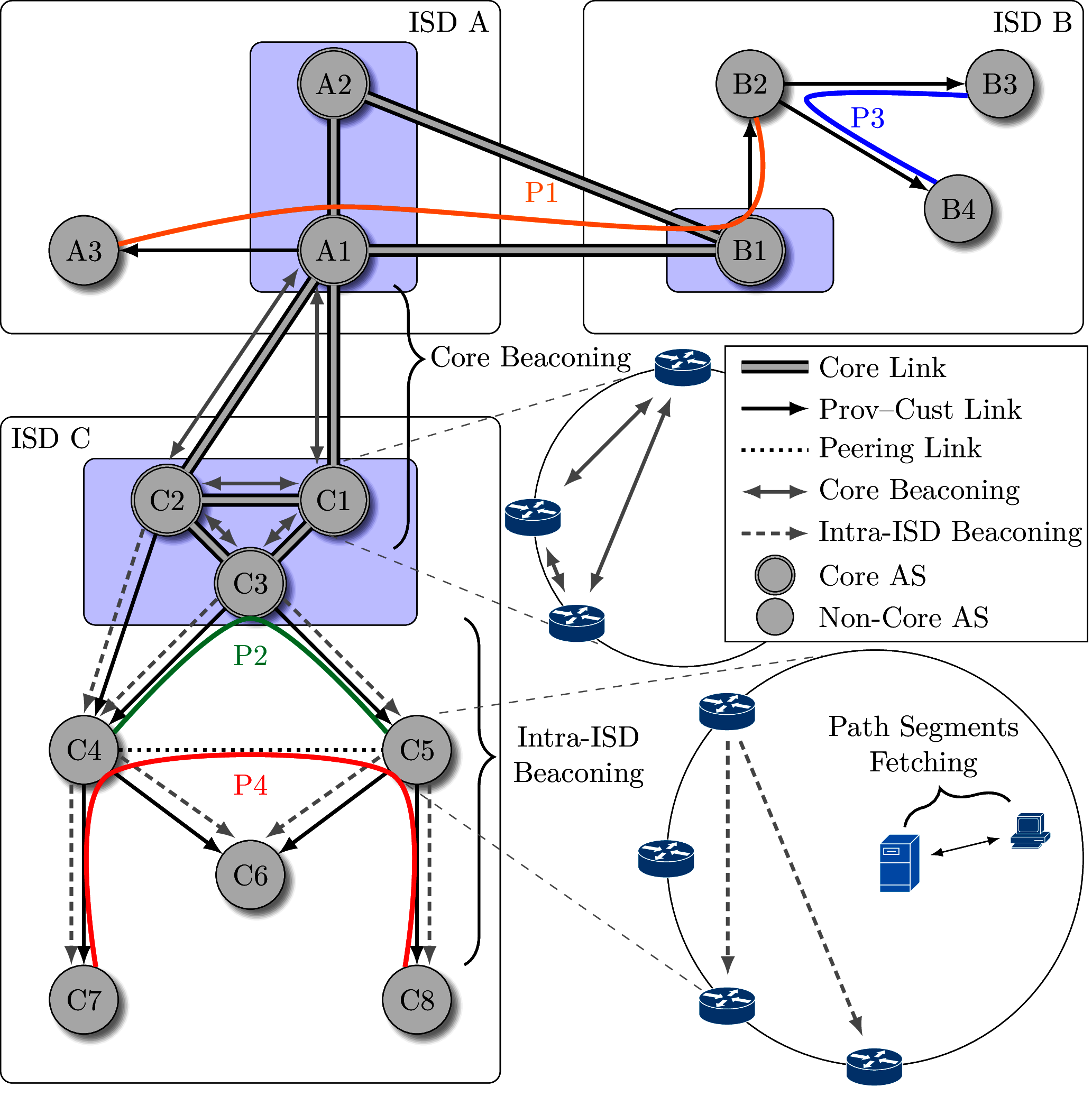}
  \caption{\label{fig:scion-overview} A SCION topology consisting of three ISDs and several possible end-to-end paths.}
\end{figure}

ASes in SCION are grouped into isolation domains (ISD) which provide routing isolation.
Within each ISD, ASes are further split into core ASes, which provide connectivity to ASes in other ISDs, and non-core ASes, which provide connectivity within the ISD\@.
\Cref{fig:scion-overview} shows a SCION network with three ISDs.
In SCION, the control plane is responsible for creating path segments, a process called ``beaconing'', (1) between core and non-core ASes within a single ISD, and (2) between core ASes of all ISDs.
These path segments ensure that the complete topology remains connected.
A path segment consists of a sequence of ASes that are traversed.
Path segments are constructed iteratively by attaching local routing information, such as the incoming and outgoing border routers, at each traversed AS\@.
These routing messages are digitally signed by each AS and contain all necessary information for an endpoint to forward traffic on this path segment.
Additionally, they can be extended with arbitrary per-AS attributes, which we leverage in \system{}.

The process of creating segments differs depending on the type of path segment created.
To construct core-path segments connecting any pair of core ASes, every core AS initiates a routing message, which is then flooded through the whole core network.
For example, C3 creates a core routing messages and sends it to both C1 and C2, which then forward it to A1, etc.
Connectivity within an ISD is established as follows:
Path segment creation is initiated by core ASes, which send routing messages to their customer ASes, which will in turn forward them on to their respective customers, until all leaf ASes are reached.
An example of such an intra-ISD path segment creation is core AS C3 creating and sending a routing message via C5 to C8 to connect C8 to the network.

\subsubsection{Data Plane}
An endpoint queries the local SCION control service for path segments to a certain destination.
These path segments are then combined into end-to-end paths.
Typically, two intra-ISD-path segments and one core-path segment is used (see P1), but some of these segments may be omitted, e.g., when communicating within an ISD (see P2 and P3) or via peering links (see P4)\@.
Each of these path segments contains a compact hop-field per traversed AS\@.
Endpoints then encode these hop fields in the headers of data packets to specify the inter-domain forwarding path at the granularity of in- and egress routers of the traversed ASes.

%% file: prediction.tex
In this section, we highlight issues prevalent in existing latency prediction systems, motivate the need of \system{} through three concrete use cases, and discuss design challenges.

\subsection{Prior Work on Latency Prediction}
Latency prediction for intra- and inter-domain networks has been extensively studied in the past.
Latency prediction approaches can be separated into deterministic approaches that distribute aggregated \emph{measured latencies}, and model-based approaches that use statistical \emph{latency models} to infer latency based on network delay monitoring between various vantage points in the network~\cite{Tabatabaeimehr2021,Perdices2019,Krasniqi2020}.
Due to challenges such as asymmetric routing, unpredictable paths, dynamic load balancing, and lack of transparency on the actually used inter-domain forwarding paths, recent advancements in latency prediction have been focused on model-based approaches.

\mysubsubsection{Intra-domain Latency Prediction}
In the space of intra-domain networking, there has been much work on low-latency networking~\cite{Li2021} and latency prediction (often in the context of data centers)~\cite{Bouzidi2018,Alesawi2019,Guo2015}.
This shows the usefulness of latency prediction in a controlled environment to enhance application performance, reduce overhead, and potentially minimize cost in terms of hardware or computational resources.

\mysubsubsection{Inter-Domain Latency Prediction}
In the space of inter-domain networking, low-latency networking and latency prediction is hindered by a multitude of factors.
Since forwarding devices are not controlled by a single entity, they may behave unexpectedly.
Additionally, external factors, such as varying queuing delay caused by background traffic, impact the prediction accuracy.
Moreover, the lack of transparency in today's Internet reveals little about the used network paths, further increasing the uncertainty of any form of latency predictions.
Nevertheless, there has been extensive research in latency prediction in inter-domain networking, demonstrating the desire for latency prediction~\cite{Tabatabaeimehr2021,Khatouni2019,Liu2015a,Perdices2019,Krasniqi2020}.

\subsection{Use Cases}\label{sec:prediction:use-cases}
With the emergence of inter-domain path-aware network architectures, some of the issues prevalent in latency prediction can be addressed.
Path-aware networks enable path transparency, which allows endpoints to know the network paths of their traffic, and consequently make use of latency-related information about these paths.
Furthermore, path-aware \emph{multipath} networks provide multiple (partially) disjoint network paths for an endpoint to choose from.
Endpoints can thus locally optimize paths based on a metric, e.g., latency.
We argue that recent developments in inter-domain path-aware networks allow users to accurately \emph{estimate} propagation delay through concrete measurements, enabling innovative and novel use cases.
Concretely, we present three use cases.

\mysubsubsection{Delay-Based Congestion Control}
The first use case is providing propagation delay information to delay-based congestion control algorithms to accurately derive the current latency inflation on a path and infer the level of congestion.
The knowledge of propagation delay is thus critical for the competition of the delay-sensitive BBR algorithm with the traditional loss-based CUBIC algorithm.
In this competition, BBR has been found to be unfair towards CUBIC in terms of capacity sharing, because the sending rate of the BBR flows is directly proportional to their estimate of the path propagation delay.
Crucially, BBR flows overestimate this propagation delay as a result of queuing caused by competing CUBIC flows, and therefore maintain an excessively large sending rate~\cite{hock2017experimental,mishra2022we}.
This unfairness can be eliminated with explicit knowledge of path propagation delay, as offered by \system{}.

\mysubsubsection{(First-Packet) Latency Estimate}\label{sec:use-cases:first-packet-latency-estimate}
The next two use ca\-ses illustrate the importance of accurate latency information in path-aware networks that offer multiple paths to choose from.
For short-lived connections, e.g., query-response protocols such as DNS, it is essential that an RTT estimate is available \emph{before} sending the first packet.
There is no benefit for an endpoint to actively measure the latency if all relevant data can be transmitted in the first reply, i.e., the reply may arrive before the measurements are finished.

\begin{figure}[t]
  \centering
  \includegraphics[width=\linewidth]{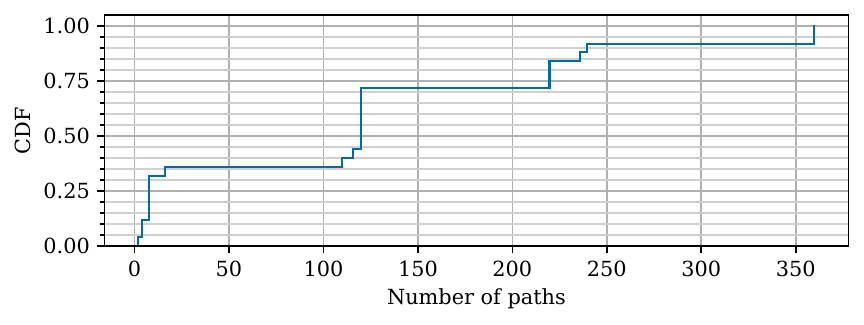}
  \caption{The number of distinct paths to all reachable ASes in the commercial SCION network from a single vantage point.}\label{fig:scion-deployment-number-of-paths}
\end{figure}

\mysubsubsection{Efficient Probing}
Finally, we argue that a propagation latency estimate is \emph{necessary} to efficiently probe a multipath network for low-latency paths.
Consider an endpoint establishing and maintaining a long-lived low-latency connection.
If only a few network paths are available, the endpoint could simply continuously measure the paths and switch to a lower latency path if available.
However, this approach does not scale if the network offers hundreds of potential paths, i.e., in a massive multipath network, as this would incur substantial processing, time, and bandwidth overhead.
The current commercial SCION network provides a concrete example of such a scenario.
\Cref{fig:scion-deployment-number-of-paths} depicts the number of different paths we can observe from ETH Zürich to 30 ASes located in 5 different ISDs.
Although the SCION network is small compared to the Internet, the median number of paths to each destination is over 100, and we expect this number to further increase with the expansion of the commercial SCION Internet.
The endpoint could heuristically select and probe a subset of paths but this does not guarantee finding latency-optimal paths.
We discuss a concrete algorithm for efficient path probing based on \system{} in \cref{sec:efficient-path-probing}.

\subsection{Challenges}\label{sec:challenges}
In this section, we list the most important challenges encountered in designing a latency transparency system.

\mysubsubsection{Variable Delays} One major challenge in latency estimation is to correctly model all variable delays in the network~\cite{Tabatabaeimehr2021}.
The dominating variable delay is typically queuing delay due to cross traffic filling up a packet queue on the path.
The packet processing and transmission delays are usually negligible in comparison to the propagation and queuing delays, but may be included in a latency measurement or calculated based on the link bandwidth.
Since the queuing delay cannot be predicted, our work focuses on the predictable part of the end-to-end latency, in particular the propagation delay.

\mysubsubsection{Disclosing Internal Topology} While entities can enhance latency estimates by revealing detailed information about their internal topologies, this might reveal sensitive information to a competitor. It is thus imperative that participating entities can decide how much information is revealed.

\mysubsubsection{Load Balancing and Variable Routes} Internet traffic is rerouted for various reasons, e.g., economic impact, SLAs, and bottlenecks.
This can happen for inter-domain paths or for intra-domain paths, e.g., due to traffic engineering or failing links.
Accurate latency estimation is challenging under these circumstances and requires regularly updated measurements.

%% file: system-overview.tex
We propose a latency transparency system that measures and distributes the propagation latency of inter-domain paths at Internet scale and present an efficient path probing algorithm for multipath networks.
In designing such a system, we make use of the facts that latency is an additive metric, and any inter-domain path can be split into several \emph{intra-domain} paths and \emph{inter-domain} links connecting them.
The propagation latency of an inter-domain path can thus be computed by accumulating the propagation latencies of all intra-domain paths and inter-domain links.
Therefore, we divide \system{} into two subsystems: (1) the \emph{measurement system} that accurately measures the latency of intra-domain paths and inter-domain links, and (2) the \emph{dissemination system} that globally disseminates latency information.

\subsection{Performing Latency Measurements}\label{sec:latency-measurements}
In \system{}, we focus on measuring propagation delay, instead of modeling queuing delay.
Note that depending on the measurement method, the propagation delay measurement may include the processing and transmission delay, but in practice, they are typically negligible in comparison to propagation and queuing delay.
Hence, \system{} requires a participating AS to be able to measure the propagation latency of intra-domain paths between their own border routers and of the links connecting them to neighboring ASes' border routers as shown in \cref{fig:segment-combination}.

There exist a wide variety of ways to measure latency, which can generally be separated into three groups~\cite{Tan2021}: (1) traditional network measurements~\cite{Sommer2002,RFC3917}, (2) SDN-based measurements~\cite{Nunes2014,Yu2015}, and (3) telemetry-based measurements~\cite{Kim2015,RFC8321}, each with different tradeoffs.
Since we are interested in the \emph{one-way propagation} latency, the network operator must ensure that queuing delay is factored out, e.g., via packet prioritization or exclusion of ``packet queue time'', and that potential path asymmetry is taken into consideration, e.g., using one-way latency measurements with synchronized clocks instead of RTT measurements.
Additionally, multiple paths with varying latency may exist between border routers due to redundancy, load balancing, or traffic engineering, and paths may change over time.
The network operator should measure all possible paths and re-measure changed paths to revoke the out-of-date path segment and re-disseminate it.
Optionally, the network operator may also enhance the measurement with the used measurement methodology and a level of measurement (un-)certainty, e.g., assigning a higher level of certainty if a more sophisticated latency measurement approach is used.

\subsection{Disseminating Latency Information}\label{sec:latency_dissemination}
To calculate the latency of an inter-domain path, the endpoints must be provided with the latencies of its constituent intra-domain paths and inter-domain links.
\system{} achieves this in a scalable fashion through path exploration and dissemination.
In the exploration phase, each AS on an inter-domain path encodes the latency information of its AS hop in the forwarded path segment.
In the dissemination phase, the endpoint retrieves the path segments with the included latency information.
Note that \system{} uses an opt-in approach and an AS can choose to disclose latency information with appropriate granularity based on the desired level of topology secrecy.
The privacy aspects are discussed in more detail in \cref{sec:discussion:deployment-model}.

\begin{figure}[t]
  \centering
  \includegraphics[height=5.8cm]{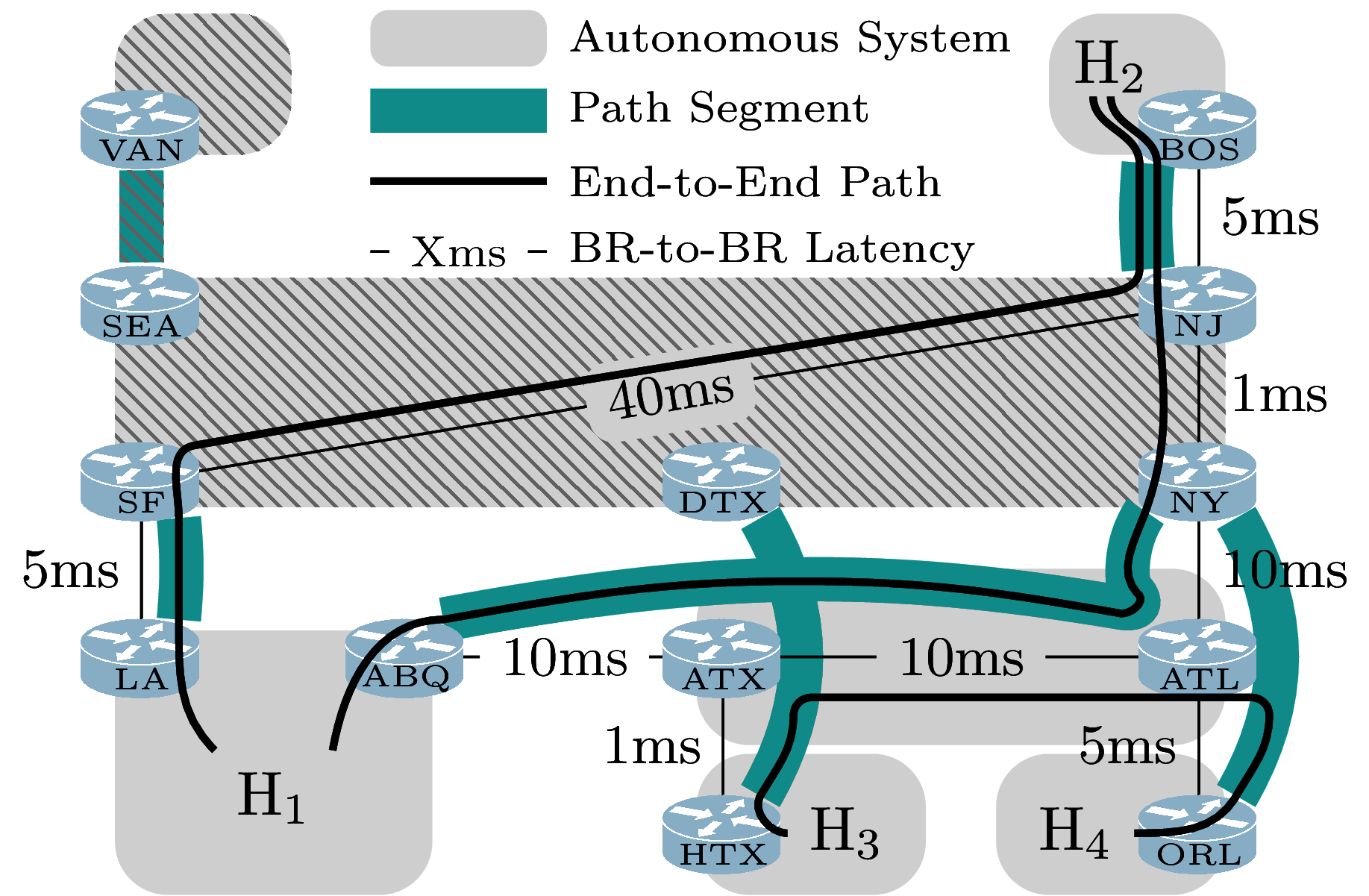}
  \caption{Topology where four hosts ($\text{H}_{1-4}$) combine path segments to end-to-end paths and leverage \system{} to select paths. Diagonal lines indicate core ASes and core path segments.}\label{fig:segment-combination}
\end{figure}

The latency information of each AS hop consists of the propagation latency of the intra-domain path between the ingress and the egress border routers (e.g., SF to NJ in \cref{fig:segment-combination}) as well as the propagation latency of the inter-domain links between the border routers of the neighboring ASes (e.g., SF to LA).
Due to possibly asymmetric latencies, the AS encodes the latency in both directions.
If the exact intra-domain path each packet can take is not predictable, e.g., due to load-balancing or backup routes, the AS must disseminate the minimum propagation latency of all possible routes---which is necessary for efficient path probing, see \cref{sec:prediction:use-cases}.
In addition to this minimum latency, an AS can optionally provide more information, such as the maximum propagation latency among these paths or a latency distribution.
Alternatively, the network operator could make such intra-domain paths (including their latency) accessible via FABRID policies~\cite{Kraehenbuehl2023} and allow endpoints to explicitly select a specific intra-domain path.

\mysubsubsection{Disseminating Latency in Hierarchical Architectures}
In hierarchical routing architectures like SCION, path segments need to be combined to construct full inter-domain paths.
Hence, not only do we need to add the latency information in each path segment, but we also need to ensure that endpoints can reconstruct the latency information of all possible segment combinations.
\Cref{fig:segment-combination} shows an example where the lack of intra-AS latency information between SF and NJ would cause $\text{H}_1$ to incorrectly select the path through LA (via SF, NJ, BOS) to communicate with $\text{H}_2$ even though the alternative path through ABQ (via ATX, ATL, NY, NJ, BOS) has a smaller propagation latency.
Hence, \system{} must disseminate the latency information between all border router pairs of ASes at the path segment junctions.
To satisfy this requirement of SCION's hierarchical routing with two levels, i.e., core and intra-ISD routing, we propose two different information dissemination mechanisms, one for each level.

In core routing, where the topology is densely connected and routing messages are flooded to a subset of neighbors, we need a mechanism with small overhead per routing message to achieve scalability.
To that end, each AS only encodes the latency for the intra- and inter-domain path that the routing message traversed.
This is possible since core-path segments are always combined with intra-ISD-path segments, which will contain the necessary latency information to interfaces that are not traversed by the core-path segment.

In intra-ISD routing, where routing messages are only forwarded from providers to customers and the topology is typically sparse, more latency information can be added to routing messages without introducing significant overhead.
A trivial approach is the addition of latency information between the egress interface of the routing message and all other interfaces.
However, this approach wastes network bandwidth, since some latency information will be duplicated, i.e., two combined path segments will both contain the latency information between two interfaces.
To prevent this, only the latency information between the egress interface and interfaces which have a lower AS-local identifier is added.
This ensures that any two intra-ISD-path segments, or any core and any intra-ISD-path segment can be combined.
For example, in \cref{fig:segment-combination}, hosts $\text{H}_3$ and $\text{H}_4$ need the intra-AS latency between ATX and ATL even though neither segment contains this link.

\mysubsubsection{Scalability}
\system{} achieves scalability using three mechanisms.
First, it disseminates the minimum propagation latency of a path, which does not change frequently, and thus does not require frequent re-dissemination.
Second, the existing path-segment establishment and dissemination process is used to disseminate the latency information without introducing additional messages.
Third, it leverages the observation that core path segments are always combined in their entirety with intra-ISD path segments, thus no additional intra-AS latency information needs to be added to the core path segments, but only to the intra-ISD path segments.

\mysubsubsection{Obtaining Full Inter-Domain Path Latency}
Once provided with the path segments to construct full inter-domain paths, endpoints compute the full path latency by accumulating latencies of constituent AS hops.
Note that an endpoint may only receive partial latency information if some on-path ASes do not reveal this information.
In such a case, the endpoint locally decides whether to use the partial information, e.g., by assuming a propagation delay of zero for the missing latency information, or to discard the incomplete latency information.

\begin{figure}[t]
  \centering
  \includegraphics[height=4.25cm]{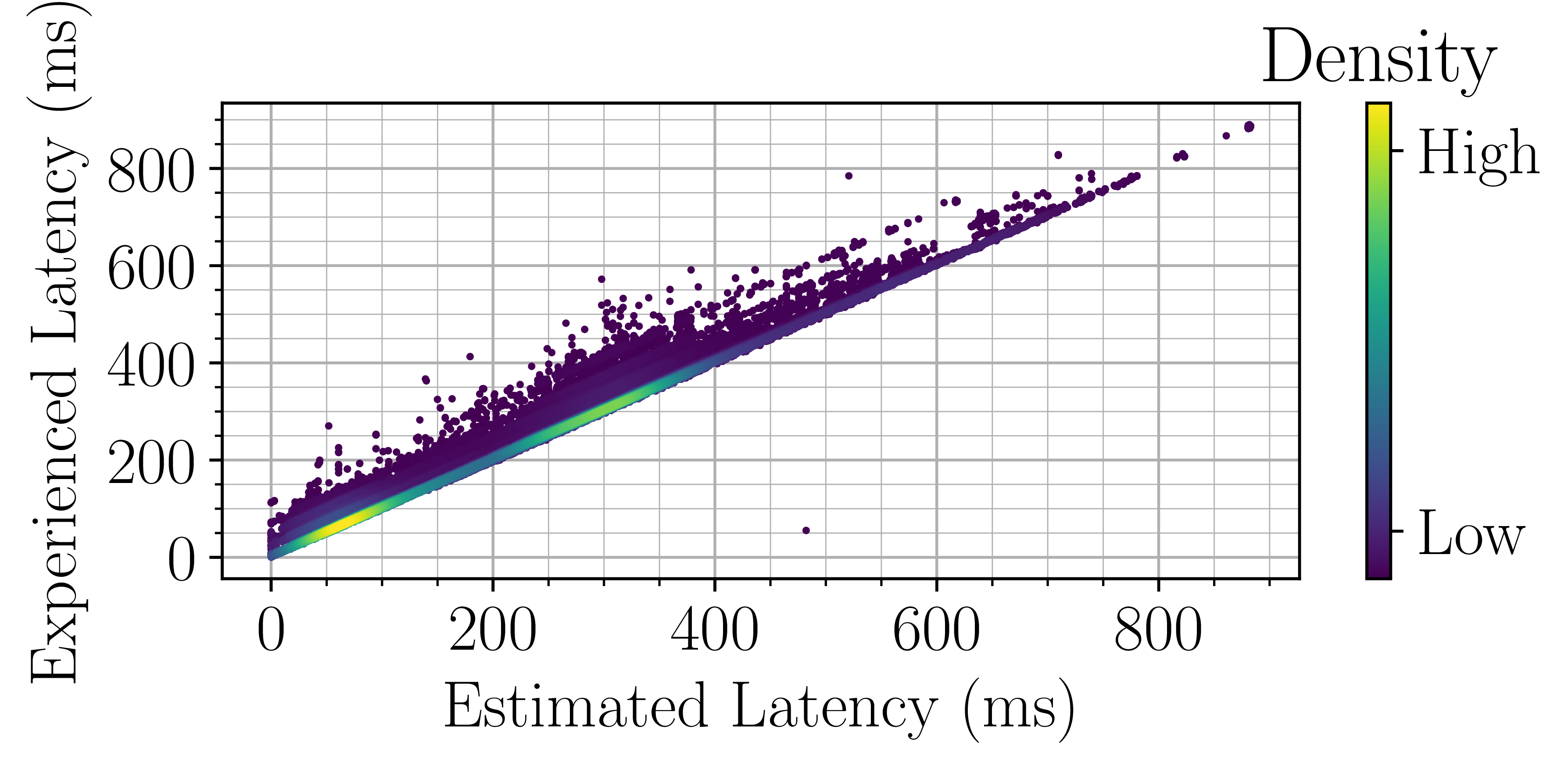}
  \caption{Difference between the estimated and the experienced propagation latency in SCIONLab.}\label{fig:scionlab-rtt-measurement-performance-scatter}
\end{figure}

\subsection{Efficient Path Probing}\label{sec:efficient-path-probing}
We propose an efficient path probing algorithm based on \system{}, which works as follows:
(1) sort all paths based on their propagation latencies (assuming zero latency for path segments with missing latency information), (2) measure them one-by-one (or batch-wise) and remember the smallest measured latency, and (3) stop as soon as the propagation delay of the next path is higher than the smallest measured latency.
Since all following paths \textit{must} have higher latency, we can terminate the algorithm early.
A requirement is that, if a path has multiple intra- or inter-domain links with different propagation delays, the system always returns the \emph{minimum} value.
Otherwise, the algorithm might discard and thus not probe a potential candidate path.
The number of probed paths depends on the distribution of latencies, e.g., a higher variance in propagation delays leads to an earlier termination.

%% file: evaluation.tex
We prove the feasibility of \system{} by implementing it on the SCIONLab testbed, evaluate the impact of propagation latency knowledge on the fairness of delay-based congestion control, and show the benefits of \system{}'s first packet latency estimation through simulation on Internet-scale topologies.

\subsection{Implementation and Deployment in SCIONLab}
We evaluate \system{} in the SCIONLab testbed~\cite{kwon2020scionlab}.
Our implementation uses RTT measurements, i.e., the minimum measured RTT over the span of \SI{60}{\second}, to infer propagation latency between AS interfaces.
The experienced latency is then measured by taking the median of 5 ping measurements.

\Cref{fig:scionlab-rtt-measurement-performance-scatter} shows that inferring propagation latency based on the RTT, which is arguably the simplest measurement approach, provides a reasonable approximation of the propagation latency since less than 5\% of measurements experienced a lower latency than the advertised propagation latency.
The overall experienced latency, with a median of \SI{234}{\milli\second} and a 95th percentile of \SI{551}{\milli\second}, is relatively high since the SCIONLab network partially consists of overlay links, which may lead to longer paths, and since the SCIONLab nodes are globally distributed in Europe, the United States, and East Asia.
However, note that SCIONLab, as a research network, experiences little congestion.
In networks carrying large amounts of data and experiencing more congestion, these results might differ substantially and require more sophisticated latency measurement systems.

\begin{figure}[t]
  \includegraphics[width=\linewidth]{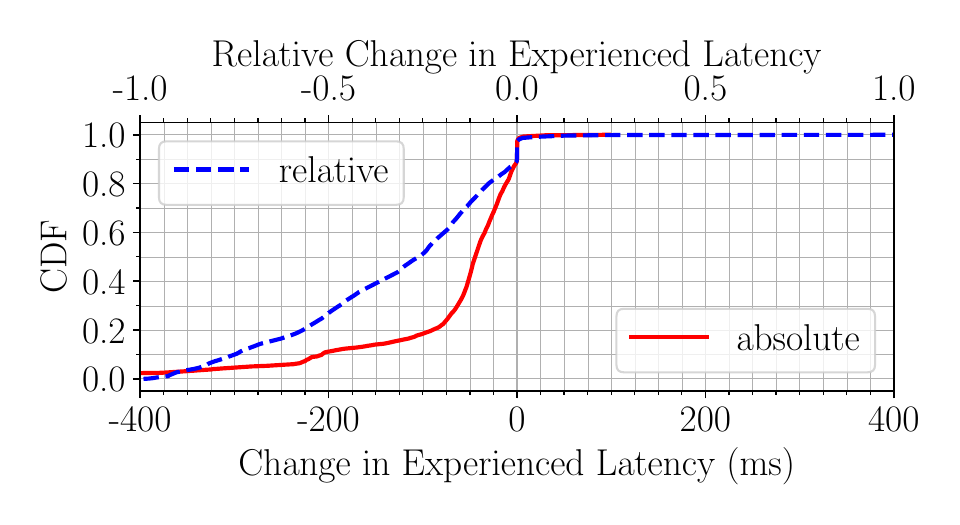}
  \caption{Latency reduction (in relative values or \si{\milli\second}) when selecting the path with the lowest propagation latency.}\label{fig:scionlab-improvement}
\end{figure}

In addition to the feasibility of the system, we measure the latency reduction of an endpoint that chooses the path with the lowest estimated propagation latency over the default path.
\Cref{fig:scionlab-improvement} shows that using \system{}, 50\% of endpoints reduced the latency by over \SI{40}{\milli\second}, or 25\% relative to the original value.
In 10\% of the cases, endpoints reduced the latency by over \SI{200}{\milli\second}, or 70\%.
These results show that a multipath-capable network with inter-domain propagation latency estimate can indeed significantly reduce the experienced latency.

\subsection{Delay-Based Congestion Control}
In the following, we demonstrate how information about path propagation
delay can improve the deployment effects of delay-based congestion-control
algorithms (CCAs). In particular, we show that the knowledge of 
path propagation delay improves the fairness of the delay-sensitive
BBR CCA~\cite{cardwell2017bbr} toward the traditional loss-based CUBIC CCA~\cite{ha2008cubic}.

\begin{figure}[t]
  \centering
	\includegraphics[width=0.9\linewidth]{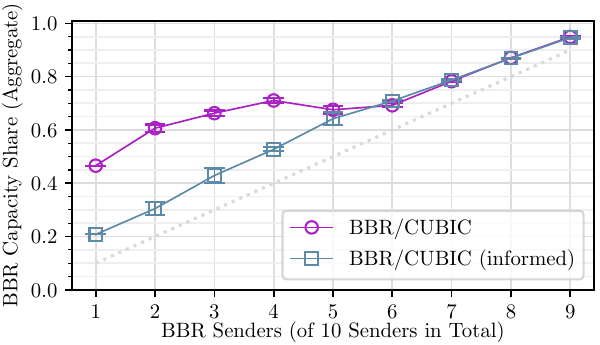}
	\caption{Bandwidth-sharing fairness between BBR and CUBIC flows, with and without explicit propagation delay knowledge.}\label{fig:evaluation:bbr-experiment}
\end{figure}

\mysubsubsection{Setup}
For our evaluation, we emulate the competition among 10 flows on a \SI{100}{Mbps} bottleneck link using Mininet~\cite{mininet_intro}.
Each flow is using a path with \SI{20}{\milli\second} one-way propagation delay,
composed of \SI{10}{\milli\second} propagation delay on non-shared links
and \SI{10}{\milli\second} propagation delay on the shared bottleneck link.
The non-shared links and the shared bottleneck link are
intermediated by a switch with a queue size that corresponds 
to 1.5 bandwidth-delay products of the network path.
In this setup, we measure the capacity share obtained by
all BBR flows together, depending on how many of the 10 flows 
adopt BBR, while the remaining flows adopt CUBIC.
We also distinguish two different BBR versions, namely
the official BBR version, and a variant named ``informed BBR'' that we implemented by adapting the BBR source code:
While the official BBR attempts to estimate path propagation delays by regular probing, informed BBR works with actual propagation delays, which are known when using \system{}.

\mysubsubsection{Results}
\Cref{fig:evaluation:bbr-experiment} presents the results
of this experiment.
For all tested configurations, the BBR capacity share is 
strictly above the dotted gray line that marks the 
proportional capacity share; hence, BBR is
unfair towards CUBIC in every evaluated case.
However, this unfairness is particularly pronounced
if fewer than 5 flows adopt the official BBR version.

Crucially, this unfairness stems from the RTT-probing
behavior of the BBR flows, which try to discover
the propagation delay by emptying on-path buffers with
sharp contractions of their sending rate, and
use the resulting measurement to adjust 
their congestion window~\cite{cardwell2017bbr}.
While this probing behavior indeed allows discovering
the propagation delay if only BBR flows share
the bottleneck link~\cite{scholz2018towards},
competing CUBIC flows preserve buffer utilization when the BBR flows
are probing, and thus inflate the propagation-delay
estimate of the BBR flows~\cite{hock2017experimental,mishra2022we}. 
As a result, the BBR flows maintain an excessively large
congestion window, resulting in an unfairly high sending rate.
Notably, this effect arises only if the CUBIC flows are
numerous enough to keep up buffer utilization in the
\SI{200}{\milli\second} time window when the BBR flows are probing; 
in our setting, fewer than 4 CUBIC flows cannot fill 
the buffer during this RTT probing, and thus do not
distort the BBR propagation-delay estimate.

Clearly, the propagation-delay estimate is never distorted for the \emph{informed} BBR version, as this version learns the actual propagation delay from \system{}.
This knowledge thus eliminates the overdimensioning of the BBR congestion window, and improves the fairness of BBR towards CUBIC.

\subsection{Large-scale Simulations}
In this section, we analyze the benefits of \system{} for first packet latency estimation, in particular DNS resolution.
We perform large-scale simulations on real-world Internet topologies from CAIDA~\cite{CAIDA-Data-Geo} to compare the performance of \system{} with BGP-based (shortest AS path) routing.
For this evaluation, we simulate BGP using SimBGP~\cite{SimBGP}, and SCION using a simulator based on ns-3~\cite{nsnam_ns-3_nodate}, to find inter-domain paths in the current Internet and in SCION+\system{}.
The latencies between routers are given by their great circle distance, calculated from their geographical location.
\Cref{sec:appendix:simulations} provides a detailed explanation of the topologies, parameters, and simulation frameworks used.

\begin{figure}[t]
  \includegraphics[width=\linewidth]{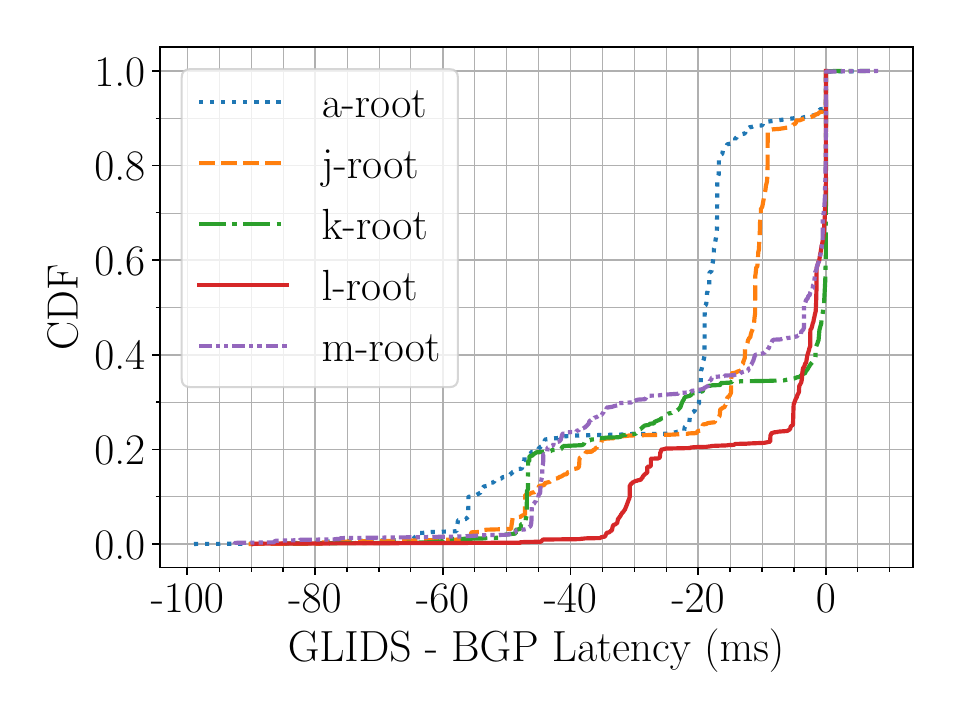}
  \caption{Comparison of lower-bound latencies from probes to DNS root servers between \system{} and BGP.}\label{fig:simulation_results}
\end{figure}

Based on these simulations, we analyze by how much \system{} could reduce this lower-bound (great circle) latency between clients and DNS root servers.
\Cref{fig:simulation_results} illustrates the difference between the simulated latency in SCION and BGP, showing that SCION's latency-aware routing and path selection can reduce the lower-bound latency to the 5 root servers in our topology for at least 40\% of probes.
Furthermore, this latency deflation is at least 20 ms for 20\% of probes to any of the root servers.
However, the latency to root servers from a negligible portion of probes can be inflated by less than 10 ms due to suboptimal choice of destination AS ingress interface.
This is caused by a lack of information on the relative position of the destination root server in the destination AS\@.

%% file: discussion.tex
This section focuses on the practical aspects of achieving latency transparency through \system{}.
In particular, we focus on \system{}'s incremental deployment model, potential hurdles for adoption, and its privacy aspects.
Furthermore, we discuss the veracity of the disseminated latency information and highlight additional use cases related to bandwidth reservation.

\subsection{Deployment Model}\label{sec:discussion:deployment-model}
A crucial aspect of any new Internet protocol is deployability~\cite{RFC8170}, i.e., to ensure compatibility with existing and legacy systems in a partial deployment while providing clear incentives for early adopters.
Ideally, existing latency measurement mechanisms and hardware can be reused for internal latency measurements, allowing \system{} to be deployed at low cost.

\system{} can be incrementally deployed by having only a subset of ASes include latency information in their routing messages.
While this does not yet provide full propagation latency transparency for all Internet users, as long as ASes that are topologically close to the communicating endpoints offer \system{}, partial latency information inferred by endpoints can be used for latency-based path optimization.
Hence, two endpoints purchasing Internet connectivity from Tier 2 ISPs can achieve propagation latency transparency if their ISPs support \system{} and are customers of the same \system{}-enabled Tier 1 AS\@.
Even if no Tier 1 AS supports \system{}, ISPs can leverage peering and transit links to other ISPs to locally achieve propagation latency transparency.

The incentive of an early \system{} adopter is to attract network traffic by providing propagation latency information.
Note that even if the early adopter does not offer better performance than its competitors, simply the presence of the latency information may lead to endpoints sending traffic on these paths to benefit from the various use cases presented in \cref{sec:prediction:use-cases}.
This holds true especially for delay-based congestion control and first-packet latency estimation, while the usefulness of efficient probing increases together with the adoption rate since more paths can be pruned.

Another important aspect of \system{} is privacy, since the participating ASes reveal propagation latencies of internal paths.
We argue that privacy is often not a concern in \system{} since it typically does not reveal more sensitive information, such as the exact configuration of internal routers and links, compared to the existing SCION paths.
Furthermore, each AS can decide not to reveal certain (privacy-sensitive) latencies.
Finally, even in today's Internet, propagation latencies can often be inferred through measurements between different vantage points (although typically with a lower precision).

\subsection{Veracity of Latency Information}
One important aspect of \system{} is ensuring the veracity of the disseminated latency information.
A dishonest AS may disseminate wrong information for a financial gain.
Artificially increasing the disseminated propagation latencies is typically less problematic since users will send their latency-sensitive traffic on alternative paths.
Artificially decreasing the disseminated propagation latencies may attract users to send their latency-sensitive traffic on a sub-optimal path.
In \system{}, the veracity of latency information is protected by including a timestamped signature from the AS that provides the latency information in the respective routing message.
Hence, \system{} provides non-re\-pudi\-ation to ensure that a misbehaving AS can be punished while preventing framing attacks.
Note that while this provides the necessary building blocks to implement a system to detect misbehaving ASes, the full design of such a system is beyond the scope of this paper.

\subsection{Opportunities For Propagation Latency Transparency}
In addition to the presented use cases, another intriguing use of propagation latency transparency is to enhance path selection for traffic that intrinsically exhibits low queuing delays and where the propagation latency is thus close to the experienced latency.
For example, constant rate traffic, which is typically observed in fixed bit rate video conferencing, may not incur significant queuing delay as long as the aggregate rate is below the minimum link capacity and packet pacing is used.
Furthermore, instead of relying on all communicating parties to send traffic at their allowed maximum rate, the allowed rate can be guaranteed through bandwidth reservation systems~\cite{Giuliari2021,Wyss2022}.
Propagation latency knowledge may then in turn allow users to select an optimal low latency path with bandwidth reservation, i.e., optimizing both for the required bandwidth and low latency.

%% file: related.tex
We separate related work into three categories: latency measurement systems, measurement-based intra-domain latency prediction, and model-based latency prediction.

\begin{sloppypar}
\subsection{Latency Measurement Systems}
Tan et al.~\cite{Tan2021} provide an extensive overview of network measurement approaches.
They divide them into three categories:
(1) Traditional network measurement using active (e.g., ping and traceroute), passive (e.g., NetFlow~\cite{Sommer2002}, sFlow~\cite{RFC3176}, or IPFIX~\cite{RFC3917}), and hybrid measurements, i.e., combination of the two, (2) Software-defined measurements based on SDN~\cite{Nunes2014} and PDP~\cite{Bifulco2018} (e.g., SLAM~\cite{Yu2015}), and (3) telemetry-based approaches, in particular in-band telemetry, which adds telemetry data to data-plane packets.
Examples of in band telemetry are: In-band Network Telemetry (INT)~\cite{Kim2015}, In-situ Operation Administration and Maintenance (IOAM)~\cite{I-D.brockners-inband-oam-requirements}, Alternate Marking-Performance Measurement (AM-PM)~\cite{RFC8321}, and Active Network Telemetry (ANT)~\cite{Pan2019}.
All of these measurement systems can be used in \system{} to measure one-way delays of both intra-domain paths and inter-domain links.
\end{sloppypar}

\subsection{Intra-Domain Latency Prediction}
Latency prediction is well researched in the setting of intra-domain networks, where high bandwidth, low latency, and many-to-one communication is common.
However, this work focuses on inter-domain latency transparency, which faces different challenges, such as the multi-domain environment, less agency over distant links and nodes, and scalable dissemination of latency information.

LACO~\cite{Zanzi2021} provides latency-driven network slicing in 5G radio networks by allowing a provider to slice its network. It requires a single provider to offer network slices to tenants, which is different from the distributed nature of \system{}.

Hermes~\cite{Zhang2017} reroutes traffic in datacenters to avoid congestion and active path probing through RTT measurements.
However, detecting congested links and routing around them in an Internet-scale topology is significantly more challenging than in datacenters with limited topology size.

Pingmesh~\cite{Guo2015} is a large-scale ping-based latency measurement system analyzing the health and performance within and between data centers of a single operator. Deploying such a system in an inter-domain network, brings additional challenges regarding trust and cooperation.

\begin{sloppypar}
\subsection{Model-Based Latency Prediction}
Model-based ap\-proaches attempt to construct a latency model for all Internet paths based on available measurements for known paths. Systems proposed by Tabatabaeimehr et al.~\cite{Tabatabaeimehr2021}, Krasniqi et al.~\cite{Krasniqi2020}, and Perdices et al.~\cite{Perdices2019} are notable examples. However, providing global latency transparency using model-based latency predictions is challenging since they not only rely on temporal but also on spatial predictions, which reduces the certainty of the provided guarantees.
  In comparison, \system{} does not require high-quality traffic matrices, measures and deterministically combines concrete latency measurements of each segment, and distributes latency information via SCION's scalable control plane.
\end{sloppypar}

%% file: conclusion.tex
\system{} offers exciting new possibilities for inter-domain networks and moves us closer to the prospect of global latency transparency.
We show that path-aware networks solve current network problems but also highlight novel challenges encountered in such networks and how to overcome them.

%% file: ack.tex
We thank the anonymous IFIP Networking conference reviewers for their insighful feedback and suggestions.
We gratefully acknowledge support for this project from the Werner Siemens Stiftung (WSS) Centre for Cyber Trust at ETH Zurich, from armasuisse Science and Technology, and from the European Union's Horizon 2020 research and innovation programme under grant agreements No 825310 and 825322.
Finally, we thank Marc Wyss, Giacomo Giuliari, and Marc Frei for their valuable feedback.

%% file: appendix.tex
\subsection{Large-scale Simulations}\label{sec:appendix:simulations}
We evaluate the impact of \system{} on latency-optimized inter-domain routing in SCION using simulations.
We compare the lower-bound latency from the location of active RIPE Atlas probes to the location of DNS root servers.
We exclude probes and root servers whose AS numbers are not included in our topology.
This setup allows us to obtain realistic and accurate locations of both the probes and the DNS root servers.

For SCION, we assume that each probe calculates the great circle latency to all egress interfaces of its own AS\@.
Then, it combines this latency with the latency of all paths from its local AS to the ASes of all root servers, and select the lowest-latency combination.
However, the probes do \textit{not} have access to the latency between the root servers and the ingress interfaces in the destination ASes.
We assume the destination AS sends the received packets on each ingress interface to the nearest root server.
In some cases, BGP selects an ingress interface that is closer to the nearest root server than the ingress interface selected by GLIDS\@.
In that case, BGP \emph{may} provide a lower-latency end-to-end path.

For BGP, we assume that each probe sends its packets to the closest border router, and all border routers forward packets to the closest next border router on the selected AS-path.
At the destination AS, the border router sends packets to the nearest root server.

\mysubsubsection{Topology.}
We use the CAIDA AS rel-geo data set~\cite{CAIDA-Data-Geo} containing the relationships between \num{12000} ASes as well as locations and number of links between neighboring ASes.
We assume that the border routers connected to a link are located in the same location as the link.
Then, we use these locations to compute great circle latencies between border routers of each AS\@.
Finally, we extract the \num{2000} highest-degree Tier-1 and Tier-2 ASes by incrementally pruning the lowest-degree ASes, and simulate BGP and SCION on the extracted topology.

\mysubsubsection{Simulation Setup.}
To simulate dissemination of latency information via routing messages, we use the SCION simulator that was developed based on the ns-3 network simulator~\cite{nsnam_ns-3_nodate}.
We modify routing logic in the simulator to encode latency information into routing messages as described in~\cref{sec:latency_dissemination}.
Moreover, we forward routing messages for low propagation latency paths with higher priority to ensure that endpoints receive paths with low propagation latency.
Since our dataset does not contain intra-domain path latencies, we approximate the propagation latency of intra-domain paths between the interfaces, i.e., border routers, of each AS using their great circle latency.
Finally, as peering links between neighboring ASes typically occur within a data center, we assume that this inter-domain latency is negligible.
To find paths in the current Internet we simulate BGP using the SimBGP simulator~\cite{SimBGP}.
Since we are investigating the latency to the DNS root servers, only ASes hosting DNS root servers announce prefixes (i.e., only the prefix of the root server), while other ASes do not announce any prefixes.
All ASes have the same path selection policy, first based on Gao-Rexford policy, then based on AS-path length, and finally based on distance to the next router.